\pdfoutput=1

\documentclass[11pt]{article}

\usepackage[final]{acl}

\usepackage{times}
\usepackage{latexsym}

\usepackage[T1]{fontenc}

\usepackage[utf8]{inputenc}

\usepackage{microtype}

\usepackage{inconsolata}

\usepackage{graphicx}

\usepackage{booktabs}
\usepackage{multirow}
\usepackage{amsmath}
\usepackage{subcaption} 
\usepackage{tabularx}  
\usepackage{array}     
\usepackage{bm}
\usepackage{enumitem}

\usepackage{graphicx}

\PassOptionsToPackage{table,xcdraw,HTML}{xcolor}
\usepackage{xcolor}

\usepackage{colortbl}
\usepackage[normalem]{ulem}
\useunder{\uline}{\ul}{}

\usepackage{algorithm}
\usepackage{algorithmicx}
\usepackage{algpseudocode}
\floatname{algorithm}{Algorithm} 

\title{GainRAG: Preference Alignment in Retrieval-Augmented Generation through Gain Signal Synthesis}

\author{Yi Jiang, Sendong Zhao\thanks{Corresponding author}, Jianbo Li, Haochun Wang, Bing Qin  \\
  Research Center for Social Computing and Interactive Robotics,\\
 Harbin Institute of Technology, China \\
  \texttt{\{yjiang,sdzhao,jbli,hcwang,qinb\}@ir.hit.edu.cn} \\
  }

\begin{document}
\maketitle
\begin{abstract}
The Retrieval-Augmented Generation (RAG) framework introduces a retrieval module to dynamically inject retrieved information into the input context of large language models (LLMs), and has demonstrated significant success in various NLP tasks. However, the current study points out that there is a preference gap between retrievers and LLMs in the RAG framework, which limit the further improvement of system performance. 
Some highly relevant passages may interfere with LLM reasoning because they contain complex or contradictory information; while some indirectly related or even inaccurate content may help LLM generate more accurate answers by providing suggestive information or logical clues. 
To solve this, we propose \textbf{GainRAG}, a novel approach that aligns the retriever’s and LLM’s preferences by defining a new metric, ``gain’’, which measure how well an input passage contributes to correct outputs.
Specifically, we propose a method to estimate these gain signals and train a middleware that aligns the preferences of the retriever and the LLM using only limited data.
In addition, we introduce a pseudo-passage strategy to mitigate degradation.
The experimental results on 6 datasets verify the effectiveness of GainRAG\footnote{The source code is publicly available at \url{https://github.com/liunian-Jay/GainRAG}}. 
\end{abstract}

\section{Introduction}
Large Language Models (LLMs)~\citep{achiam2023gpt,touvron2023LLaMA} perform well in processing natural language tasks, but their knowledge is fixed in model parameters and is difficult to update dynamically over time~\citep{ji2023survey,he2022rethinking}. To tackle this issue, the Retrieval Augmented Generation (RAG) framework adds a retrieval module that brings in relevant external knowledge and integrates it into the input context of the LLMs. This approach has shown impressive results across various natural language processing tasks~\citep{gao2023retrieval,lewis2020retrieval}. 
Previous work is devoted to solving two problems, namely, \textit{retrieving more relevant information} in retrieval and \textit{effectively utilizing context to generate the correct answer} in generation. 
However, they fail to address the preference gap between the retriever and the LLMs. Recent studies have highlighted that while retrieved passages may be relevant, they are not necessarily preferred for generation. 
In other words, only passages that align with the LLM's preferences can provide meaningful gain and enhance generation performance. 

\begin{figure}[p!t]
    \centering
    \includegraphics[width=0.999\linewidth]{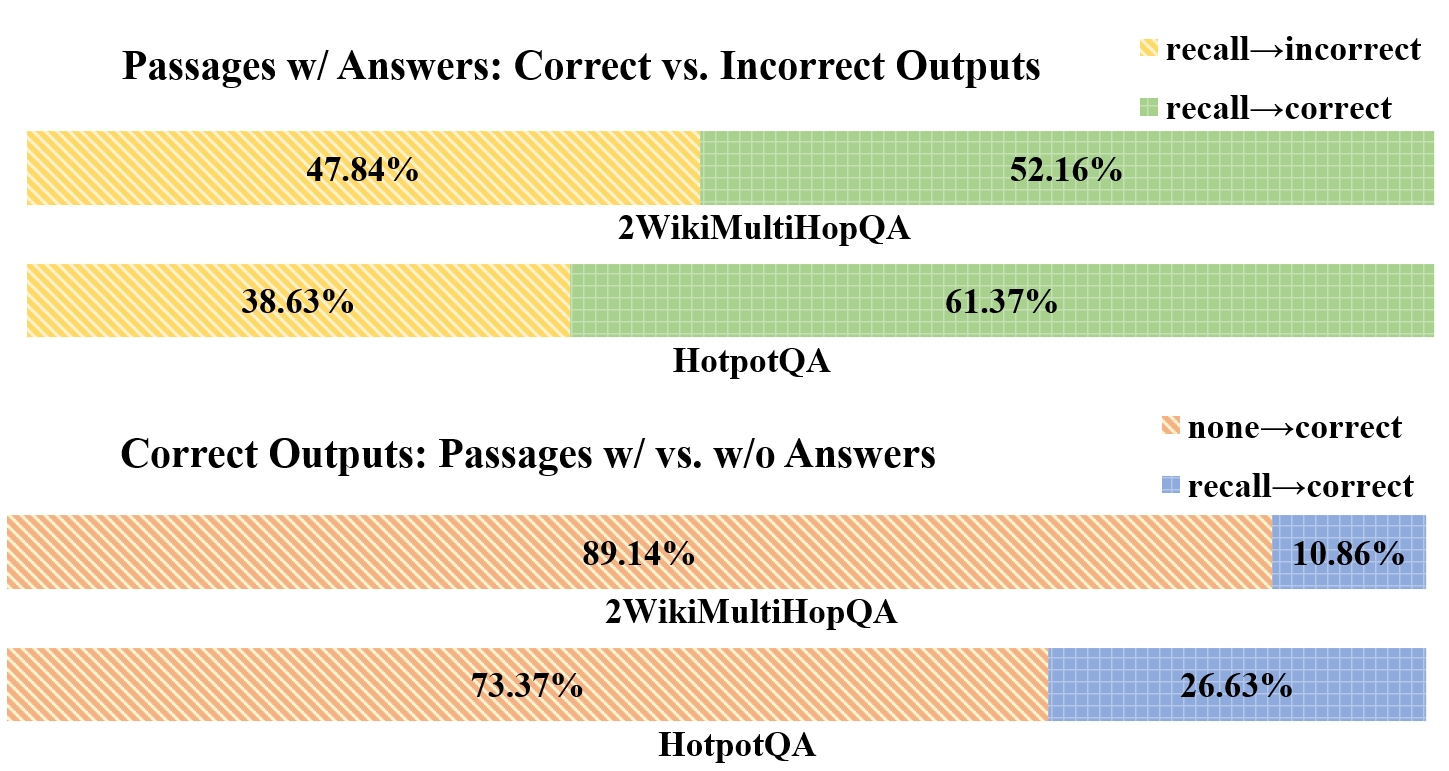}
    \caption{We analyze the preference gap between retrieved passages and LLMs on 2 datasets: HotpotQA and 2Wiki2MultiHopQA. The top shows the proportion of correct and incorrect generations when the retrieved passage contains the gold answer. The bottom shows the proportion of whether the passage used contains the golden answer when the LLM response is correct.}
    \label{fig:challenge}
\end{figure}

\begin{figure*}[!t]
    \centering
    \includegraphics[width=0.999\textwidth]{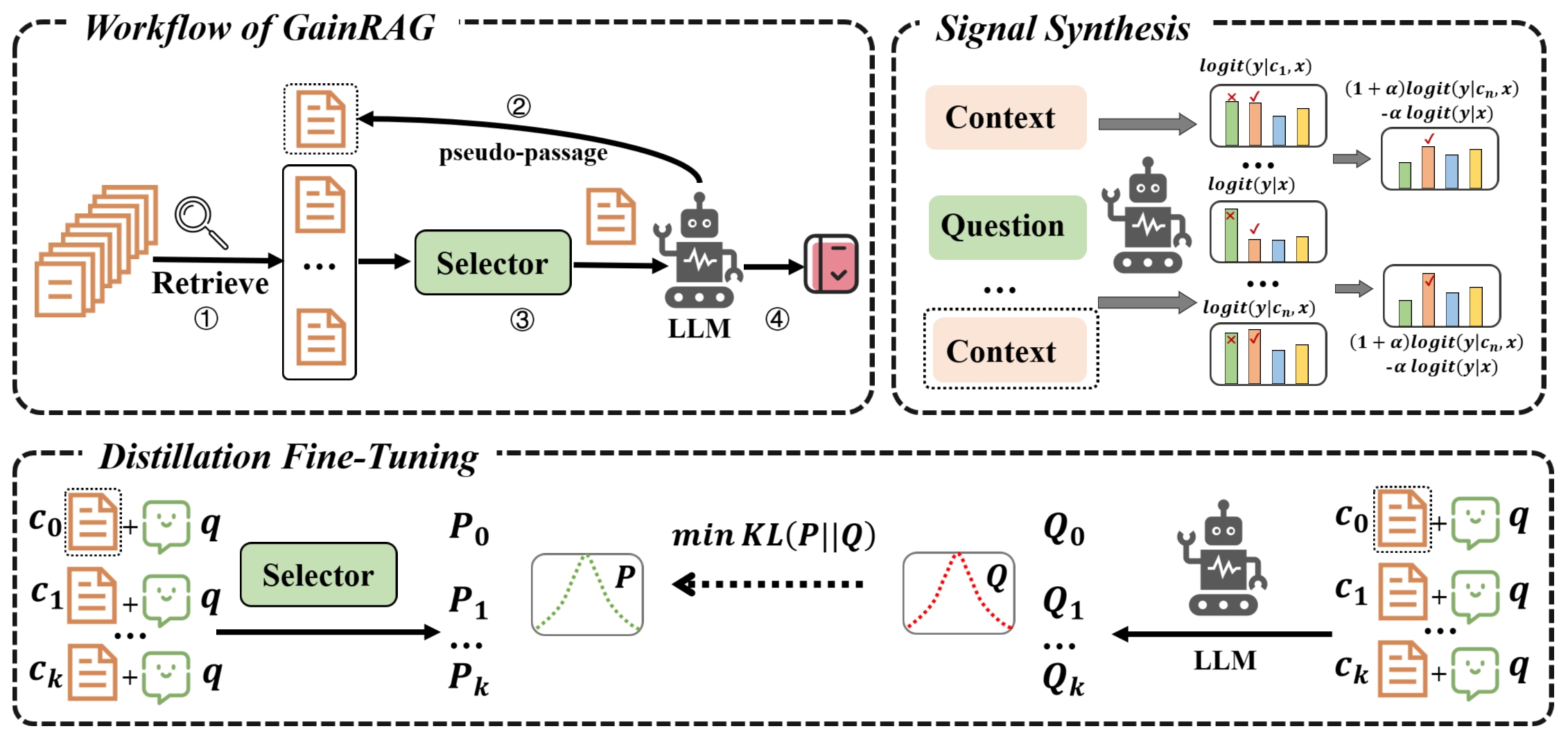}
    \caption{Illustration of the GainRAG framework. The GainRAG workflow, preference signal synthesis, and selector distillation fine-tuning are shown respectively. }
    \label{fig:framework}
\end{figure*}
Specifically, existing retrievers are usually designed based on human-defined relevance criteria, such as whether a passage directly contains the answer to a question ~\citep{ke2024bridging}. However, this approach does not fully align with the way LLMs process information. Some highly relevant passages may actually disrupt LLMs reasoning by introducing complexity or contradictions, while some seemingly unrelated or even partially incorrect content can help by offering useful hints or logical cues ~\citep{dong2024understand, cuconasu2024power}. Therefore, retrieval should shift its focus from traditional “relevance” to “gain”—prioritizing information that helps LLMs generate more accurate results.

To examine and validate this preference gap, we retrieve 100 relevant passages for each sample on two multi-hop question-answering datasets. Each passage is used to enhance the sample query for evaluation. 
As shown at the top of Fig.~\ref{fig:challenge}, 
we find that even when the retrieved passage contains the correct answer, nearly half of the samples still generate incorrect responses, indicating that while these passages are relevant, they are not particularly beneficial for generation.
As shown at the bottom of Fig.~\ref{fig:challenge}, in most cases where the generation is correct, the passages used do not directly contain the answer. 
Instead, some passages that indirectly provide answers or clues may be less relevant but more beneficial, as they align better with the LLM's preferences.

Existing work aligns the retriever with the LLM's preferences mainly by fine-tuning the retriever or training both together. 
For example, Replug~\citep{shi2023replug} aligns the retriever with LLM preferences by training the retriever directly, while RA-DIT~\citep{lin2023ra} uses dual training. However, this approach requires a large amount of high-quality data and is difficult to implement in real-world industrial settings. 
There are also some training middleware to align language model preferences, such as BGM~\citep{ke2024bridging} and DPA-RAG~\citep{dong2024understand}, but their perception and measurement of LLM preferences are coarse, capturing only basic patterns without a detailed understanding of nuanced differences.

To address the above challenges, we introduce a middleware between the retriever and the LLM to solve the problem of inconsistent preferences between the two, that is, the most profitable passages can be selected from the large number of passages retrieved by the retriever. 
Specifically, we introduce a method to quantify LLM's preference based on based on perplexity and contrastive decoding. This enables passage gain calculation and mitigates the LLM's overconfidence bias.  
Secondly, we use this to synthesize a small number of samples to distill the preference perception ability to the selector. 
Finally, we introduce a pseudo-passage strategy to prevent situations where all retrieved passages are not profitable. Together with the selector, it mitigates degradation and enables efficient integration of internal and external knowledge. 

In general, our contributions can be summarized as follows:
\begin{itemize}
    \item We analyze the preference gap between retrievers and LLMs, quantifying the LLM's preference for passages by defining ``gain’’ and introducing GainRAG to address this gap. 
    \item We provide a selector and introduce a pseudo-passage strategy that work together to not only avoid degeneration and triviality, but also achieve efficient integration of internal and external knowledge. 

    \item We train with very few samples and validate on 6 datasets. The results show excellent performance and generalization of GainRAG.
\end{itemize}

\section{Related Works}
\subsection{Retrieval-augmented Generation}
In recent years, in order to solve the problems of outdated knowledge in the model and hallucination of large language models, retrieval-augmented generation has been introduced~\citep{fan2024survey,gao2023retrieval}, and many efforts have been made in two aspects: ``how to retrieve more relevant information’’ including retriever fine-tuning~\citep{nian2024w} and query optimization~\citep{ma2023query,wang2023query2doc} and ``how to better use the retrieved information to generate answers’’ including special fine-tuning~\citep{wang2024rear,zhang2024raft} and decoding strategies~\citep{shi2023trusting}.

\subsection{Retriever-LLM Preference Alignment}
The misalignment between the preferences of the retriever and LLM results in retrieved information being ``relevant’’ but not always ``useful’’.
There are many existing works that have made efforts to solve this problem. Replug~\citep{shi2023replug} and Altas~\citep{izacard2022few} use LLM to supervise and guide the fine-tuning of the retriever. RA-DIT~\citep{lin2023ra} uses dual training allows the language model to guide the training of the retriever and the language model to adapt to the retrieved information.
However, these methods usually require a large amount of data, which makes the resources expensive and huge.
DPA-RAG~\citep{dong2024understand} synthesizes data and trains the reranker and LLM to align the preferences of the two. BGM~\citep{ke2024bridging} trains an intermediate to complete the rearrangement and selection of information from the coarsely ranked retriever. These methods simply do not clearly define and quantify this preference, which may lead to deviations in alignment. Different from them, we propose a method to quantify the preference, so that a small amount of data can be synthesized through LLM to fine-tune the selector and achieve alignment between the two.

\subsection{Bridge Between Retriever and LLM}
Many existing works achieve RAG optimization by providing a bridge between the retriever and LLM. Rerankers~\citep{glass2022re2g} such as BGE-reranker~\citep{xiao2024c} are the most common middleware, which can find more accurate passages from the information recalled by the retriever. Compressors such as RECOMP~\citep{xu2024recomp} and rewriters~\citep{ma2023query,wang2023query2doc} are also commonly used bridges, which compress the retrieved passages to achieve denoising and solve the problem of context overload. Similar to them, GainRAG we proposed introduces a preference selector to select the optimal passage, which achieves score perception and avoids excessively long context.

\section{Methodology}
In this section, we introduce GainRAG (Fig.~\ref{fig:framework}), starting with the synthesis and motivation of the gain signal, followed by selector training, and concluding with the GainRAG workflow.

\subsection{Preliminaries}
Before discussing GainRAG, we first provide a formal definition of the RAG problem.
Given a query $q$ and a corpus of documents $\mathcal{D}$,
RAG systems typically follow a retrieval-then-reading framework. 
In this approach, the retriever selects relevant passages $C = \{c_1, c_2, · · · , c_k\} \subset \mathcal{D}$  from the entire corpus $\mathcal{D}$, and the generator model (LLM) utilizes these passages $C$ to generate the answer 
$\hat{a}$. 
This process can be represented as:
\begin{equation}
    \begin{split}
        C &= \mathcal{R}(q, \mathcal{D}, k),\\
        \hat{a} &= \mathcal{G}(\mathcal{P}(q, C)),
    \end{split}
\end{equation}
where $\mathcal{R}$ denotes the retrieval function that selects $k$ relevant passages, $\mathcal{P}$ represents the prompt template that combines $q$ and $C$, and $\mathcal{G}$ is the generator, i.e., the LLM, which generates the final answer $\hat{a}$.

\subsection{Gain signal to quantify preference}
To make the passage used meet the preferences of the LLM, we need to quantify the benefit of a passage for the LLM to answer a question.
To achieve this goal, we introduce contrastive decoding~\citep{li2022contrastive} and calculate the perplexity~\citep{li2023quantity} after contrastive decoding. 

\paragraph{Perplexity} 
In instruction tuning, the model is trained to maximize the likelihood of a response given the instruction. Thus, perplexity (PPL) serves as an indicator of difficulty~\citep{li2023quantity,li2024superfiltering}.
Specifically, the PPL of a given sample $(q, a)$ is defined as
\begin{equation}
\small
    \text{PPL}(a \mid q) = \exp\left(-\frac{1}{N} \sum_{j=1}^N \log p(a_{j} \mid q, a_{1}, \dots, a_{j-1})\right).
\end{equation}
Similarly, we can also use perplexity to measure how difficult it is for the LLM to generate the correct answer $a$ for question $q$ given $c$ in RAG. 
\begin{equation}
  \small
  \text{PPL}(a \mid q, c) = \exp\!\Biggl(\!
    -\frac{1}{N} \sum_{j=1}^N 
      \log p\bigl(a_{j} \mid q, c, a_{1}, \dots, a_{j-1}\bigr)
  \!\Biggr).
\end{equation}

\paragraph{Contrastive Perplexity} 
The indicator $\text{PPL}(a\mid q,c)$ indicates the difficulty of the LLM to generate answers through query $q$ and context $c$, but it cannot distinguish whether the answer generation is driven by context $c$ or the LLM's internal knowledge. Therefore, directly using PPL for passage screening is biased. 
Inspired by CAD~\citep{shi2023trusting}, we introduce contrastive perplexity, which calculates perplexity using contrastive decoded logits~\citep{li2022contrastive} to measure the gain of context $c$ for answering $a$. This mitigates bias from the model's internal knowledge, enabling a more accurate quantification of the gain from context $c$ to query $q$. 
Given our modeling of the internal prior $p(a_t\mid q, a_{<t})$, the probability distribution of the LLM output is adjusted as: 
\begin{equation*}
\begin{aligned}
a_t &\sim \tilde{p}(a_t \mid c, q, a_{<t}) \\
&\propto p(a_t \mid c, q, a_{<t}) \left( \frac{p(a_t \mid c, q, a_{<t})}{p(a_t \mid q, a_{<t})} \right)^{\alpha}, 
\end{aligned}
\end{equation*}
where $p$ is the original probability distribution from the LLM, $\tilde{p}$ is the distribution adjusted via contrastive decoding, $a_{<t}$ is the generated sequence, and $\alpha$ is a hyperparameter controlling the degree of adjustment. 
After rearranging the formula,
\begin{equation}
\begin{aligned}
a_t \sim \text{softmax} &\big [ 
     (1 + \alpha) \, \text{logit}_{\theta}(a_t \mid c, q, a_{<t}) \\
    & - \alpha \, \text{logit}_{\theta}(a_t \mid q, a_{<t}) 
\big]. 
\end{aligned}
\end{equation}
Therefore, the gain of passage $c$ on the correct answer $a$ generated by the LLM for question $q$ can be quantified as $\mathcal{M}(c, a\mid q)$, formally, 
\begin{equation}
\small
    \mathcal{M}(c, a\mid q) = \exp\left(-\frac{1}{N} \sum_{j=1}^N 
    \log \tilde{p}\big(a_{j} \mid q, c, a_{1}, \dots, a_{j-1}\big)\right). 
\end{equation}

\subsection{Selector between LLM and Retriever}

To effectively align the retriever's output with with the preferences of the LLM, we introduce a middleware, the selector, to identify the most beneficial passages.
This selector leverages the context-gain-aware approach to refine passage selection, ensuring that the information fed into the LLM maximally enhances its performance.


The selector is formulated as a gain estimation problem, where gain estimates are distilled into a learnable function $f(q,c;\theta) \rightarrow \hat{v}$, where $\theta$ is the trainable parameter of the model. We employ the BGE pre-trained model as the foundation for the selector, defining its function mapping as:
\begin{align}
\bm{\hat{V}} &= \left[ f(q, c_0), \dots, f(q, c_k) \right],
\end{align}
where $\hat{V}$ denotes the predicted gain values for the retrieved passages $C$ with respect to query $q$. This formulation enables the selector to effectively rank passages, ensuring that the most relevant and gainful content is utilized by the LLM.

\subsection{Pseudo-passage Strategy}
While increasing the number of retrieved passages raises the likelihood of finding useful information, passage selection can still become degenerate, that is, 
all retrieved passages do not provide any useful information, or augmenting with each retrieved passage may be worse than LLM direct response. 

To address this challenge, we introduce the pseudo-passage strategy.
Concretely, before selecting any external passages, we generate a pseudo-passage $c_0$  by prompting the LLM with the query. Formally, we define: 
\begin{equation}
    c_0 = \mathcal{G}(\mathcal{P}_0(q)),  
\end{equation}
where $\mathcal{P}_0$ refers to the prompt template used to generate the pseudo-passage.
This pseudo-passage $c_0$ is then added to the selector’s candidate list alongside the truly retrieved passages. 

This strategy reduces over-reliance on potentially unhelpful retrieved passages and ensures that the selection of passages is always gain-oriented. Consequently, the pseudo-passage strategy not only mitigates degradation but also promotes the collaborative and efficient integration of internal and external knowledge of the LLM, ultimately leading to more robust performance.

\subsection{Training of the LLM Preference Selector}
As shown in Fig.~\ref{fig:framework}, we train a middleware, i.e., selector, to align the preferences of the LLM by selecting the most beneficial passage. 

\paragraph{Data Construction} Starting with a set of QA pairs, for each $\{q, a\}$ pair, we first retrieve $k$ relevant passages $C = \{c_1,\dots,c_k\}$, and then calculate the gain to construct the training set $\{(q_i,a_i,c_i^j,v_i^j) \mid v_i^j = M(c_i^j,a_i \mid q_i)\}$. 
Additionally, to mitigate degradation, we generate and add a pseudo-passage $c_0$ to the set. 
Algorithm \ref{algr-Construction} summarizes this process. 

\begin{algorithm}[!b]
\caption{Gain Signal Construction}
\label{algr-Construction}
\textbf{Input:} Original Dataset $D_o = \{(q, a), \dots \}$, Corpus $\mathcal{D}$ \\
\textbf{Output:} Enriched dataset $D_v = \{(q, c, v), \dots \}$
    \begin{algorithmic}[1]
    \State Initialize $D_v \gets \emptyset$ \Comment{Initialize}
    \For{$(q, a) \in D_o$}
        \State $c_0 \gets \mathcal{G}(\mathcal{P}_0(q))$ \Comment{Generate pseudo-passage}
        \State $[c_1, \dots, c_k] \gets \mathcal{R}(q, \mathcal{D}, k)$ \Comment{Retrieve relevant passages}

        \For{$c \in \{c_i \mid i=0 \dots k \}$}
            \State Compute $v = \mathcal{M}(c, a \mid q)$ 
            \State Add $(q, c, v)$ to $D_v$
        \EndFor
    \EndFor
    \State \textbf{return} $D_v$ \Comment{Return the enriched dataset}
    \end{algorithmic}
\end{algorithm}

\paragraph{Training Loss} 
To make the selector aware of the relative gains of different $c$ on the same $q$, we refer to ~\citet{shi2023replug,lin2023ra} and use distillation loss. 
Note that due to the long-tail distribution of $V$ and the large values at the tail, the label $V$ we actually use is a simple transformation, that is, $v=-\log(v+1)$.
The distillation loss is calculated as follows, 
\begin{equation}
\begin{aligned}
    \bm P &= \text{softmax}(V), \quad \bm V = [v_1, \dots, v_k], \\
    \bm Q &= \text{softmax}(\hat{V}), \quad \bm{\hat{V}} = [\hat{v}_1, \dots, \hat{v}_k], \\
    \mathcal{L} &= \text{KL}(\bm P \parallel \bm Q) = \sum_{i} \bm P_i \log \frac{\bm P_i}{\bm Q_i}.
\end{aligned}
\end{equation} 

\subsection{GainRAG Inference Workflow}
After obtaining the selector, it acts as a middleware to align preferences between the retriever and the LLM. 

Specifically, when a query $q$ comes, we first prompt the LLM to generate the internal information about this query and use the retriever to retrieve several relevant passages. Formally, 
\begin{align}
    c_0 &= \mathcal{G}(q), \\
    [c_1, \dots, c_k] &= \mathcal{R}(q, \mathcal{D}, k), 
\end{align}

After getting all the passages, we use the selector to predict the gain of these passages for $q$ and select the passage with the highest gain. Formally, 
\begin{align} 
    c^* &= \arg\max_{c \in \{c_0, c_1, \dots, c_k\}} f(q, c) 
\end{align}

Finally, we use the selected optimal passage for enhanced generation to obtain the predicted answer. In terms of formula, 
\begin{align}
    \hat{a} &= \mathcal{G}(q,c^*).
\end{align}
Algorithm \ref{algr-workflow} summarizes this process.

\begin{algorithm}[!t]
\caption{GainRAG Inference Workflow}
\label{algr-workflow}
\textbf{Input:} Query $q$, Corpus $\mathcal{D}$ \\
\textbf{Output:} Answer $\hat{a}$
    \begin{algorithmic}[1]
    \State $c_0 \gets \mathcal{G}(\mathcal{P}_0(q))$ \Comment{Generate pseudo-passage}
    \State $[c_1, \dots, c_k] \gets \mathcal{R}(q, \mathcal{D}, k)$ 
    \State $\hat{V} \gets []$ \Comment{Initialize score list}
    \For{$i \gets 0$ to $k$} \Comment{Iterate through all passages}
        \State $\hat{V}[i] \gets f(q, c_i)$ \Comment{Compute score for $c_i$}
    \EndFor
    
    \State $\hat{i}^* \gets \arg\max \hat{V}$
    \State $c^* \gets c_{\hat{i}^*}$ \Comment{Select the best passage $c^*$}
    \State $\hat{a} \gets \mathcal{G}(q, c^*)$ \Comment{Predict the answer}
    \State \textbf{return} $\hat{a}$ \Comment{Return the final answer}
    \end{algorithmic}
\end{algorithm}

\begin{table*}[!htbp]
\centering
\resizebox{0.77\linewidth}{!}{%
\begin{tabular}{cccccccccc}
\toprule
                         & \multicolumn{3}{c}{HotpotQA} & \multicolumn{3}{c}{2WikiMultiHopQA} & \multicolumn{3}{c}{WebQuestions} \\
\multirow{-2}{*}{Method} & EM       & F1      & Avg     & EM         & F1         & Avg       & EM        & F1        & Avg      \\
\hline
\multicolumn{10}{c}{\cellcolor[HTML]{EFEFEF}w/o retrieval}                                                                       \\
Naive                    & 22.40    & 22.44   & 22.42   & 26.80      & 20.44      & 23.62     & {\ul 44.39}     & {\ul35.90}     & {\ul40.14}    \\
GenRead                  & 31.00    & 30.50   & 30.75   & 30.60      & {\ul 25.24}      & {\ul 27.92}     & \textbf{47.69}     & 31.42     & 39.55    \\
\multicolumn{10}{c}{\cellcolor[HTML]{EFEFEF}w/ retrieval}                                                                         \\
Standard RAG             & 31.80    & 33.23   & 32.51   & 23.40      & 21.81      & 22.61     & 35.04     & 33.26     & 34.15    \\
Self-RAG                 & 30.60    & 18.83   & 24.71   & \textbf{34.00}      & 17.33      & 25.67     & 42.18     & 23.14     & 32.66    \\
Rerank             & {\ul 35.80}    & {\ul 37.45}   & {\ul 36.62}   & 24.20      & 22.94      & 23.57     & 37.50     & 35.55     & 36.52    \\
GainRAG                  & \textbf{39.60}    & \textbf{41.99}   & \textbf{40.79}   &  {\ul 31.40}      & \textbf{28.92}      & \textbf{30.16}     &  42.51      & \textbf{39.17}     & \textbf{40.84}  \\
\bottomrule
\end{tabular}%
}
\caption{EM/F1/Avg(EM,F1) of different methods experimented on datasets HotpotQA, 2WikiMultiHopQA, WebQuestions. The best and second best scores are highlighted in \textbf{bold} and \underline{underlined}, respectively.}
\label{tab:comparative-1}
\end{table*}

\begin{table*}[!h]
\centering
\resizebox{0.77\textwidth}{!}{%
\begin{tabular}{cccccccccc}
\toprule
                         & \multicolumn{3}{c}{SQuAD} & \multicolumn{3}{c}{NaturalQA} & \multicolumn{3}{c}{TriviaQA} \\
\multirow{-2}{*}{Method} & EM      & F1     & Avg    & EM       & F1       & Avg     & EM       & F1      & Avg     \\
\hline
\multicolumn{10}{c}{\cellcolor[HTML]{EFEFEF}w/o retrieval}                                                          \\
Naive                    & 18.50   & 21.57  & 20.03  & 31.25    & 29.02    & 30.13   & 60.20    & 59.96   & 60.08   \\
GenRead                  & 21.13   & 20.90  & 21.01  & {\ul 38.48}    & 32.77    & 35.62   & 64.15    & 58.94   & 61.55   \\
\multicolumn{10}{c}{\cellcolor[HTML]{EFEFEF}w/ retrieval}                                                            \\
Standard RAG             & {\ul 29.53}   & {\ul 32.46}  & {\ul 30.99}  & 38.14    & 36.82    & 37.48   & 62.16    & 61.87   & 62.02   \\
Self-RAG                 & 27.69   & 14.78  & 21.23  & 35.60    & {\ul 39.78}    & {\ul 37.69}   & 61.65    & 35.21   & 48.43   \\
Rerank            & 29.36   & 31.84  & 30.60  & 30.86    & 30.60    & 30.73   & {\ul 65.55}    & {\ul 65.09}   & {\ul 65.32}   \\
GainRAG                  & \textbf{34.65}   & \textbf{37.55}  & \textbf{36.10}  & \textbf{41.97}    & \textbf{41.27}    & \textbf{41.62}   & \textbf{67.29}    & \textbf{66.73}   & \textbf{67.01}  \\
\bottomrule
\end{tabular}%
}
\caption{EM/F1/Avg(EM,F1) of different methods experimented on datasets SQuAD, NaturalQA, TriviaQA. The best and second best scores are highlighted in \textbf{bold} and \underline{underlined}, respectively.}
\label{tab:comparative-2}
\end{table*}

\section{Experiments}
In this section, we report our experimental details and results, and provide an experimental analysis of GainRAG.

\subsection{Implementation Details}
\noindent \textbf{Training Data} 
We randomly selecte 20k samples from the HotpotQA training set and about 4k from the WebQuestions training set. For each sample, we gathere 21 relevant passages: 20 retrieved using the most common retriever Contriever~\citep{izacard2021unsupervised} and 1 generated internally.
We then applie Algorithm~\ref{algr-Construction} to add relevant passages and filter out samples where the passage with the highest gain is incorrectly generated, resulting in about 10k samples.
The decoding $\alpha$ is set to 0.5 according to CAD~\citep{shi2023trusting}.

\noindent \textbf{Training Details} 
We use LLama3-8b to generate preference values and BGE-reranker-base~\citep{xiao2024c} for the selector's initial weights, training for 2 epochs. 
All experiments are conducted on a single A100 with 80G memory.

\noindent \textbf{Inference Details}
During inference, we use Contriever~\citep{izacard2021unsupervised} as the retriever and LLama3-8B as the generator. We set the initial retrieval setting k to 100 because 100 has a high coverage, as shown in Appendix~\ref{sec:appendix-coverage} analysis experiment. For all datasets, we use 21M English Wikipedia~\citep{karpukhin2020dense} dump as the source passages for the retrieval. Prompts for the experiments can be found in Appendix~\ref{sec:appendix-prompt}

\begin{table}[!ht]
\centering
\resizebox{0.85\linewidth}{!}{%
\begin{tabular}{@{}ccc@{}}
\toprule
Task Type             & Datasets      & \# Samples \\ 
\hline
\multirow{2}{*}{Multi-HopQA} & 2WikiMultiHopQA & 500    \\
                             & HotpotQA        & 500    \\
\midrule
\multirow{3}{*}{OpenQA}      & WebQuestions    & 2032    \\
                             & NaturalQA       & 3610    \\
                             & SQuAD           & 10570   \\
                             & TriviaQA       & 11313    \\

\bottomrule
\end{tabular}
}
\caption{Description of tasks and evaluation datasets.}
\label{tab:datasets}
\end{table}

\begin{table*}[h!t]
\resizebox{0.95\textwidth}{!}{%
\begin{tabular}{lcccccccccccc}
\toprule
\multirow{2}{*}{Method} & \multicolumn{3}{c}{HotpotQA}                     & \multicolumn{3}{c}{2WikiMultiHopQA}              & \multicolumn{3}{c}{WebQuestions}                 & \multicolumn{3}{c}{NaturalQA}                    \\
                        & EM             & F1             & Avg            & EM             & F1             & Avg            & EM             & F1             & Avg            & EM             & F1             & Avg            \\
\hline
Standard RAG            & 31.80          & 33.23          & 32.51          & 23.40          & 21.81          & 22.61          & 35.04          & 33.26          & 34.15          & 38.14          & 36.82          & 37.48          \\
w/o all                 & 35.80          & 37.45          & 36.62          & 24.20          & 22.94          & 23.57          & 37.50          & 35.55          & 36.52          & 30.86          & 30.60          & 30.73          \\
w/o pseudo              & {\ul 37.80}          & {\ul 40.65}          & {\ul 39.23}          & 27.20          & 24.88          & 26.04          & 41.24          & {\ul 38.97}          & {\ul 40.11}          & {\ul 41.25}          & {\ul 40.94}    & {\ul 41.09}          \\
w/o distillation            & 34.20          & 35.85          & 35.02          & {\ul 29.60}    & {\ul 26.68}          & {\ul 28.14}          & \textbf{43.21} & 36.66          & 39.94          & 32.41          & 31.39          & 31.90          \\

GainRAG                 & \textbf{39.60} & \textbf{41.99} & \textbf{40.79} & \textbf{31.40} & \textbf{28.92} & \textbf{30.16} & {\ul 42.51}    & \textbf{39.17} & \textbf{40.84} & \textbf{41.97} & \textbf{41.27} & \textbf{41.62} \\
\bottomrule
\end{tabular}%
}
\caption{Ablation studies, including: w/o all (removing all modules i.e., the ordinary reranker), w/o pseudo (removing the strategy for generating pseudo-passage, w/o distillation (removing the distillation fine-tuning.)}
\label{tab:ablation-study}
\end{table*}

\subsection{Datasets and Evaluation Metrics}
\textbf{Eval Datasets} To verify the effectiveness and generalization of GainRAG, we use the open domain question answering datasets WebQuestion~\citep{berant2013semantic}, NaturalQA~\citep{kwiatkowski2019natural}, TriviaQA~\citep{joshi2017triviaqa} and SQuAD~\citep{rajpurkar2016squad}, as well as the complex multi-hop question answering datasets HotpotQA~\citep{yang2018hotpotqa} and 2WikiMultiHopQA~\citep{ho2020constructing}. The statistics are shown in Table~\ref{tab:datasets}. Its detailed description can be found in Appendix~\ref{sec:appendix-dataset}.

\noindent \textbf{Evaluation Metrics} We calculate exact match (EM) and F1 scores. Following ~\citet{asai2023self,mallen2022not}, we apply a non-strict \textbf{EM} metric, which considers a model's generation correct if it includes the golden answer, rather than requiring an exact match. 
F1 measures the overlap between the predicted and golden answers. 
Note that in our study, longer responses tend to increase \textbf{EM} scores due to higher matching probabilities, but often lower \textbf{F1} scores due to irrelevant content. Therefore, the average of both metrics may be a more balanced evaluation.

\subsection{Baselines}
We selected several of the most common methods for comparison.
1) \textbf{StandardRAG}, which is the most classic ``retrieve-then-read’’ paradigm.
2) \textbf{GenRead}~\citep{yu2022generate}: 
Its retriever can be seen as itself since it uses self-generated context to answer questions. It has almost no preference for misalignment, but there may be insufficient information. 
3) \textbf{Self-RAG}~\citep{asai2023self}: Through adaptive retrieval and self-criticism, it alleviates the preference misalignment problem to a certain extent.
4) \textbf{Rerank}~\citep{glass2022re2g,xiao2024c}: It is a supplement to the classic RAG. It adds middleware between the retriever and the LLM. Following the ``retrieve-rerank-read’’, we use the BGE-Reranker-base model.
For fairness, Rerank has the same settings as ours. StandardRAG and Self-RAG also only use top-1. The rest of the settings follow the settings of their original papers.

\subsection{Main Results}
Experimental results are presented in Table~\ref{tab:comparative-1} and Table~\ref{tab:comparative-2}, and we can get the following analysis:

1) Our method achieves state-of-the-art performance on almost all datasets. 
Despite using only a small subset of HotpotQA and WebQuestions for data synthesis, it generalizes well across datasets, demonstrating the robustness of GainRAG.

2) In WebQuestions, RAG methods generally underperform compared to those without external knowledge, suggesting that retrieval is not always beneficial. In cases of preference misalignment, retrieved passages can even be harmful. However, our approach still achieves the best average performance.

3) The reranking method outperforms other baselines, proving that middleware integration is both simple and effective. By leveraging a small amount of data to mitigate preference misalignment, our method significantly surpasses standard reranker.

\subsection{Ablation Study}
In order to verify the effectiveness of each module, we conducted ablation experiments on several datasets. 
The results, shown in Table~\ref{tab:ablation-study}, confirm that every module plays a crucial and irreplaceable role. The key findings are:

1) Without distillation fine-tuning, performance drops significantly, highlighting the importance of preference alignment. 
However, due to the existence of the pseudo-passage strategy, the performance can still be significantly improved over the common methods.

2) When pseudo passage strategy is absent, performance drops on some datasets significantly but not all. It shows that it plays a significant role when there is a degenerate solution, that is, when the correct preferred passage cannot be retrieved. 


3) Pseudo passages and preference perception fine-tuning are complementary and essential. Together, they prevent degenerate solutions and improve passage selection, aligning preferences between the retriever and the LLM.

\begin{figure}[!t]
    \centering
    \includegraphics[width=0.99\linewidth]{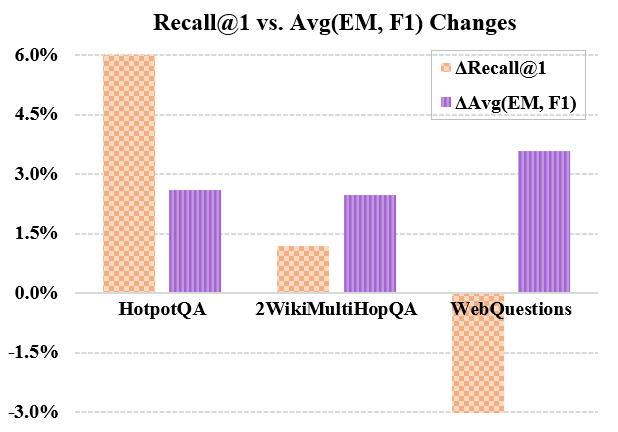}
    \caption{Illustration of gain. Changes in recall of the gold answer and downstream performance after using GainRAG.}
    \label{fig:gain}
\end{figure} 
\subsection{Effect of Preference Selection}
In order to explore why GainRAG improves the response performance of downstream LLMs, we removed the pseudo-passage strategy and calculated the changes in Recall@1 and downstream generation metrics.  

As shown in Fig.~\ref{fig:gain}, we find that there are three general cases. 
1) Our selector sometimes significantly improves the Recall@1, which further improves downstream performance. 
2) Our selector does not significantly improve the Recall@1, but the downstream performance are significantly improved. 
3) Our selector reduces the Recall@1, but the downstream performance are significantly improved. 

Case 1 is intuitive, while Cases 2 and 3 demonstrate that our selector’s impact goes beyond just relevance, highlighting the benefits of selection based on gain. This further confirms the effectiveness of our approach.

\begin{figure}[!h]
    \centering
    \includegraphics[width=1.02\linewidth]{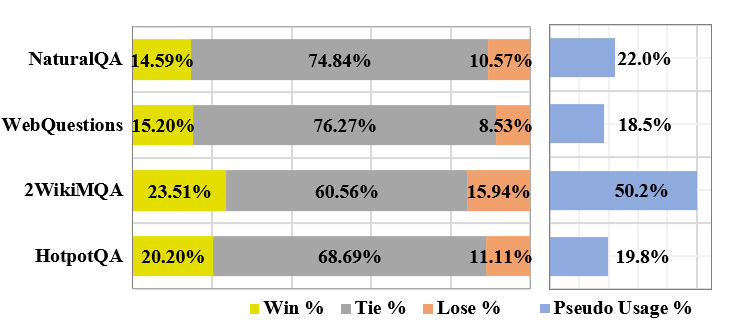}
    \caption{Illustration of the pseudo-passages generated for each dataset to avoid degenerate solutions.}
    \label{fig:pseudo-passages}
\end{figure}
\subsection{What is the Effect of Pseudo-passage?} 
To examine the role of pseudo-passages in mitigating performance degradation, we analyze their usage across different datasets.  
Specifically, we count the overall use of pseudo-passages, as shown in the right part of Fig.~\ref{fig:pseudo-passages}. In addition, we replace these cases with non-pseudo-passage with the largest gain and performed a Win-Tie-Lose comparison, as shown in the left part of Fig.~\ref{fig:pseudo-passages}.

In many cases, internally generated passages are selected, with internal knowledge used in 50\% of the cases on 2WikiMultiHopQA. This is consistent with the comparative experiment, which shows that GenRead significantly improves performance on 2WikiMultiHopQA, highlighting the value of LLM-generated passages for this dataset.
Additionally, the Win-Tie-Lose comparison reveals that the number of winning cases after replacing pseudo-passages with the highest-gain passages far outweighs the losing cases, further demonstrating the effectiveness of pseudo-passages in alleviating degradation.

\begin{figure}[!hb]
    \centering
    \includegraphics[width=1.0\linewidth]{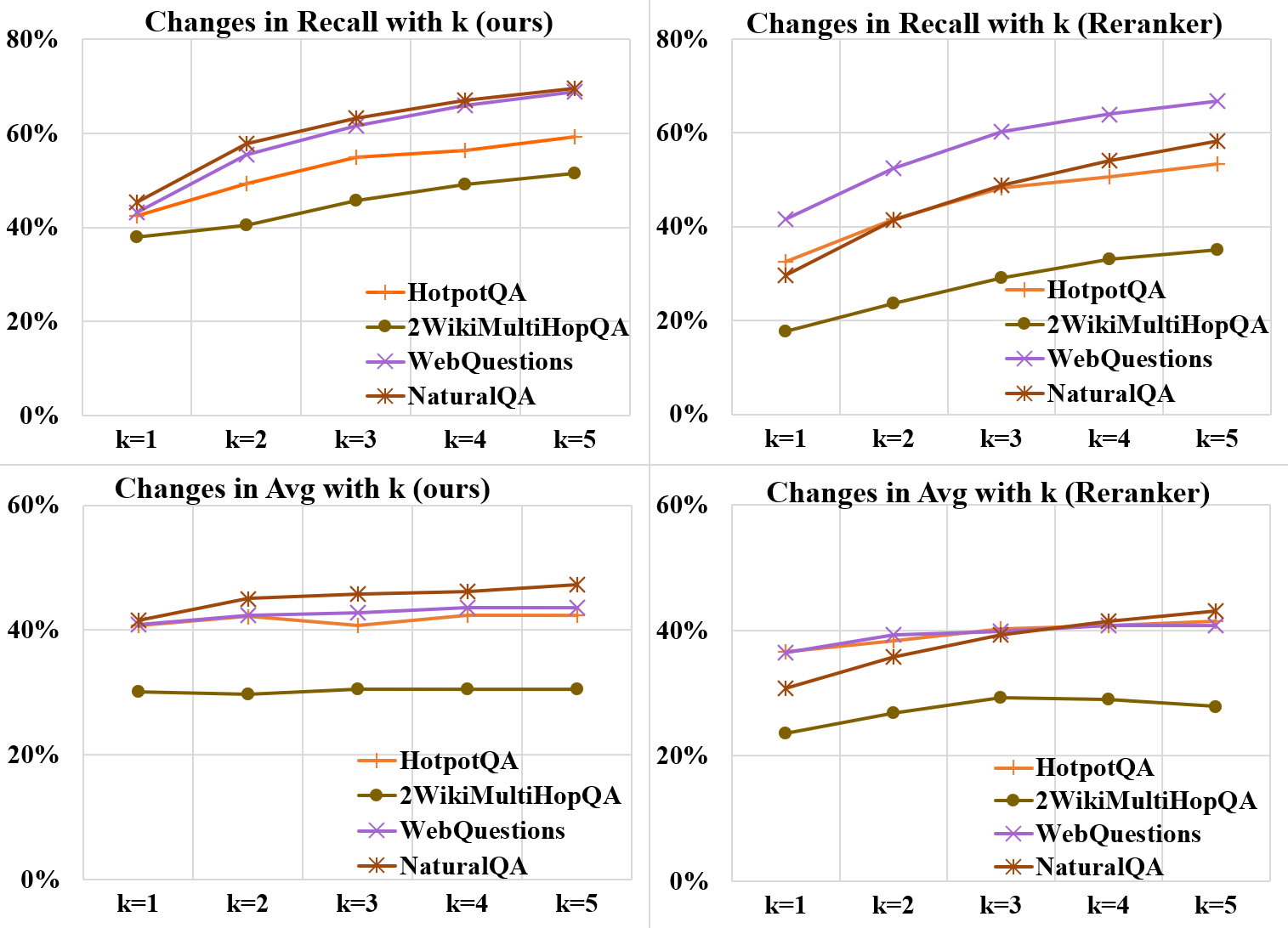}
    \caption{As the number of passages increases, the changes in recall and downstream generation performance. The left part is the change of our selector, the right side is the BGE-reranker, and the upper and lower parts are recall and Avg(EM, F1) respectively. }
    \label{fig:k}
\end{figure}

\subsection{Synthetic Signal Analysis}
To explore the effect of contrastive decoding debiasing, we use the ordinary PPL synthetic signal to fine-tune the selector under the same settings. Its performance and changes are shown in Table~\ref{tab:contrast}.

We find that contrastive debiasing is crucial, as removing it weakens preference perception for individual passages. This is because this strategy enhances the model’s perception of gain rather than just relevance, thereby alleviating over-reliance on LLM internal knowledge. 

\begin{table}[!h]
\resizebox{\linewidth}{!}{%
\begin{tabular}{cc}
\toprule
Datasets        & EM \textbf{/} F1 \textbf{/} Avg(EM,F1)        \\
\midrule
HotpotQA & 38.2 ($\downarrow$ 1.40) \textbf{/} 41.38 ($\downarrow$ 0.61) \textbf{/} 39.79 ($\downarrow$ 1.00) \\
2WikiMQA & 29.4 ($\downarrow$ 2.00) \textbf{/} 27.12 ($\downarrow$ 1.80) \textbf{/} 28.26 ($\downarrow$ 1.90) \\
\bottomrule
\end{tabular}%
}
\caption{Performance degradation after removing contrastive decoding}
\label{tab:contrast}
\end{table}

\subsection{In-Depth Comparison with the Reranker}
To explore the impact of the number of selector choices on performance and compare it with the reranker, we set the selector to the interval [1,5] and observe the performance changes, as shown in Fig.~\ref{fig:k}.
The results reveal the following:

1) Increasing K significantly boosts the recall rate, as expected, since more passages increase the likelihood of including the gold answer.

2) For our selector, selecting the top passage is usually sufficient. Even for general rerankers, increasing K does not improve downstream generation performance, and longer contexts even add overhead and may be harmful. 

3) While the recall rate increases, downstream generation performance remains largely unchanged, highlighting the scientificity and rationality of selection. This further supports the observation that simply including the gold answer does not guarantee correct generation.

\section{Conclusion}
This work analyzes the preference gap between retrievers and LLMs and proposes GainRAG to address this misalignment. We define and quantify preferences, then fine-tune a selector with signals from a small number of samples. By adding a selector and using a pseudo-passage strategy to prevent degradation, GainRAG effectively integrates internal and external knowledge of LLMs, achieving superior performance.


\section*{Acknowledgements}
This work was supported in part by National Natural Science Foundation of China [62206079]; and the Heilongjiang Provincial Natural Science Foundation of China [2023ZX01A11]. We also appreciate the support from China Mobile Group Heilongjiang Co., Ltd. @ on our research, the research is jointly completed by both parties. 

\section*{Limitations}
GainRAG selects passages with gain by calculating the gain score. However, this selection may not be the optimal solution. Whether there are some combinations of passages that make the gain stronger remains to be verified. And we only used a very small amount of training data to show the effect. In the future, large-scale data training experiments are still needed to verify whether it will get better performance. In addition, for the signal generation of large-scale data, whether a small model can be used as a generator when generating signals to accelerate the experiment is also a need for further experimental verification.


\bibliography{custom}


\appendix
\section{Dataset}
\label{sec:appendix-dataset}
Here, we introduce in detail the datasets we used, which are seven datasets on four tasks.

\textbf{2WikiMultiHopQA}~\citep{ho2020constructing} and \textbf{HotpotQA}~\citep{yang2018hotpotqa}: Both datasets are multi-hop question answering datasets based on Wikipedia. Considering the limitation of experimental cost, we used the sub-sampling set published by \citet{trivedi2022interleaving,kim2024sure}, which is obtained by extracting 500 questions from the validation set of each dataset.

\textbf{WebQuestions}~\citep{berant2013semantic}: Constructed from questions posed by the Google Suggest API, where the answers are specific entities listed in Freebase.

\textbf{NaturalQA}~\citep{kwiatkowski2019natural}: A dataset designed to support comprehensive QA systems. It consists of questions from real Google search queries. The corresponding answers are text spans from Wikipedia articles, carefully identified by human annotators.

\textbf{SQuAD}~\citep{rajpurkar2016squad}: It is a dataset for evaluating reading comprehension, created by annotators who generate questions based on the documents they read. It is widely used for training and testing open-domain QA systems.

\textbf{TriviaQA}~\citep{joshi2017triviaqa}: A compilation of trivia questions paired with answers, both originally pulled from online sources.

\section{Coverage Study}
\label{sec:appendix-coverage}
To analyze the performance of the original retriever, we conducted experiments on four datasets. Specifically, we used the retriever to retrieve [1, 5, 10, 20, 50, 100] paragraphs respectively and calculated the recall, EM coverage, and F1 coverage. For EM coverage, we used each paragraph to enhance the query separately, and as long as there is one correct response, it is considered to be covered. For EM coverage, we used each paragraph to enhance the query separately, took the response with the largest F1 value, and calculated the average of the overall dataset.

\begin{figure}[!h]
    \centering
    \includegraphics[width=0.875\linewidth]{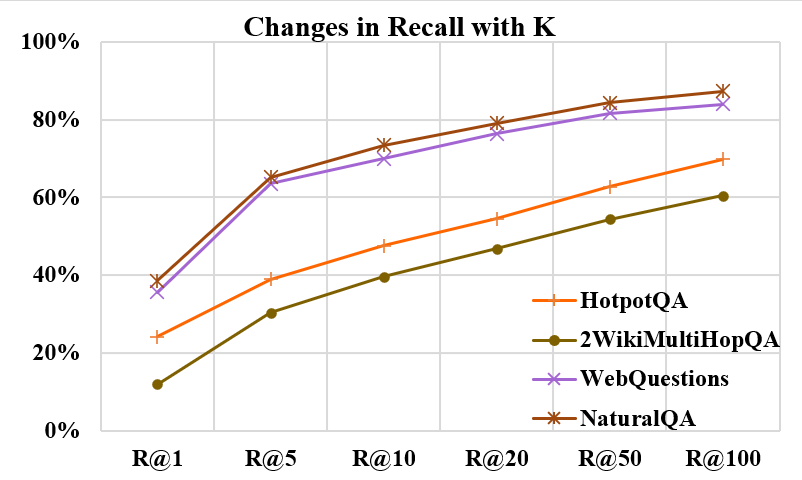}
    \caption{Illustration of the change in recall as the number of retrievals K increases}
    \label{fig:coverage-recall}
\end{figure}

\begin{figure}[!h]
    \centering
    \includegraphics[width=0.875\linewidth]{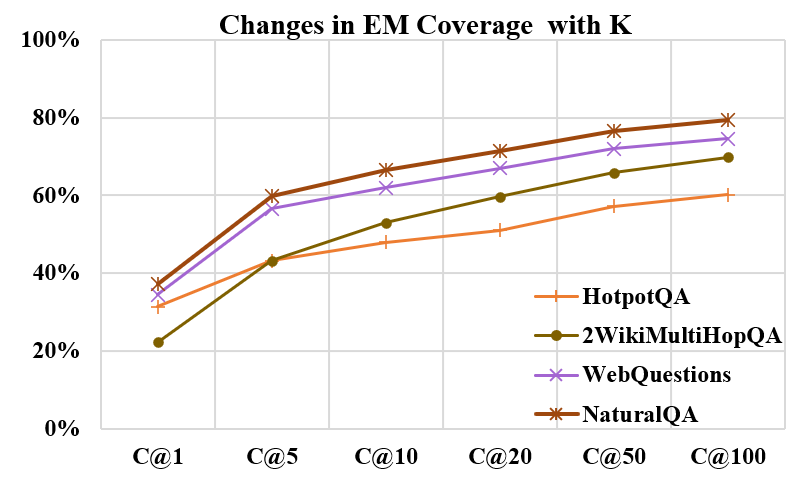}
    \caption{Illustration of the change in EM coverage as the number of retrievals K increases}
    \label{fig:coverage-em}
\end{figure}

\begin{figure}[!h]
    \centering
    \includegraphics[width=0.875\linewidth]{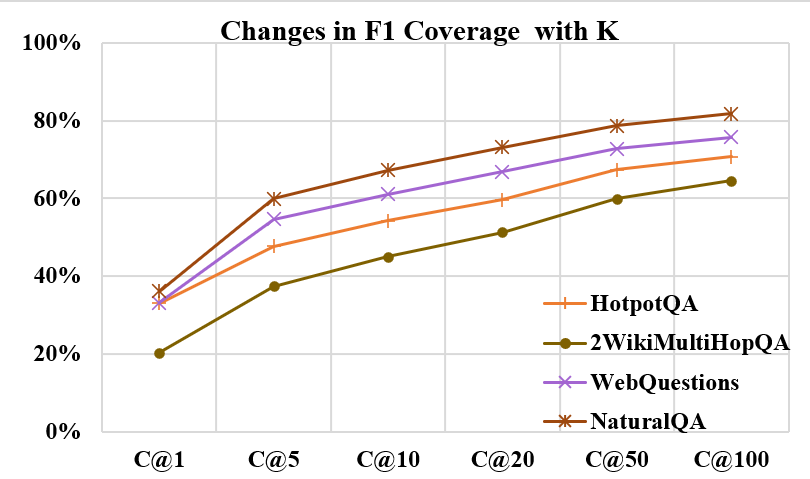}
    \caption{Illustration of the change in F1 coverage as the number of retrievals K increases}
    \label{fig:coverage-f1}
\end{figure}

As shown in Fig.~\ref{fig:coverage-recall},Fig.~\ref{fig:coverage-em} and Fig.~\ref{fig:coverage-f1}, as K increases, the recall and coverage will steadily increase. When retrieving 100, the coverage is large enough and far exceeds that of the current state-of-the-art RAG method.

\section{Training Details}
\label{sec:appendix-detail}
We use LLama3-8b as our generator for generating preference values and BGE-reranker-base~\citep{xiao2024c} for initializing the selector. We train selection for 2 epochs. 
During fine-tuning, we set train-group-size to 16 and batch-size to 8.
In addition, during training, we randomly select 16 out of 21 passages to ensure generalization. The rest of the settings follow the official fine-tuning script~\citep{xiao2024c}. 
Regarding the training data, we randomly selected 20,000 samples from the Hotpot training set and all 3,778 samples from the WebQuestion training set. After filtering, we finally obtained 14,084 training data. 
All experiments are conducted on a single A100 with 80G memory.

\section{Prompt Templates}
\label{sec:appendix-prompt}
All the prompt templates used by our proposed GainRAG are shown in Table~\ref{tab:prompt-task}.

\begin{table}[h!t]
    \centering
    \begin{tabularx}{\linewidth}{lX}
        \toprule
        \textbf{Task}        & \textbf{Task Instruction} \\
        \hline 
        Generation       & \{$passage$\} {\textbackslash n} \#\#\# Instruction: {\textbackslash n} Answer the question below concisely in a few words. {\textbackslash n\textbackslash n} \#\#\# Input: {\textbackslash n} \{$query$\}   \\
        Pseudo-Passage   & Please provide background for the question below in 100 words. Do not respond with anything other than background. If you do not know or are unsure, please generate ``N/A’’ directly. Question: \{$query$\} \\
        \bottomrule
    \end{tabularx}
    \caption{Full list of instructions used during zero-shot evaluations and pseudo-passage generation. Where $query$ and $passage$ are the paragraph to be used and the question to be answered.}
    \label{tab:prompt-task}
\end{table}

\end{document}